\documentclass[journal]{IEEEtran}
\usepackage{blindtext}
\usepackage{graphicx}
\IEEEoverridecommandlockouts

\usepackage{cite}

\ifCLASSINFOpdf
\else
\fi

\usepackage{caption}

\usepackage{braket}
\usepackage{color}
\usepackage{arydshln}
\usepackage{amsmath}
\usepackage{color}
\usepackage{pifont}
\usepackage{subfig}
\usepackage{psfrag} 
\usepackage{url}
\usepackage{listings}
\usepackage{lmodern}

\graphicspath{ {images/} }
\definecolor{orange}{RGB}{255,127,0}
\newcommand{\rec}[0]{\textcolor{blue}{$\boldsymbol{+}$}}
\newcommand{\dia}[0]{\textcolor{blue}{$\boldsymbol{\times}$}}
\newcommand{\diaone}[0]{$\nearrow$}

\newcommand{\recnot}[0]{$\rightarrow$}

\newcommand{\latinword}[1]{\textsf{\itshape #1}}%

\begin{document}
%
\title{Modelling and Analysis of Quantum Key Distribution Protocols, BB84 and B92, in Communicating Quantum Processes(CQP) language and Analysing in PRISM}
%
%
%


\author{\IEEEauthorblockN{Satya Kuppam}\\
\IEEEauthorblockA{Dhirubhai Ambani Institute of Information and Communication Technology\\
Gandhinagar, India\\
Email: 201101014@daiict.ac.in, anuroopkuppam@gmail.com}
\thanks{This is a pre-print version of a journal submission, calling for comments from the audience.}
}

\maketitle

\begin{abstract}
Proof of security of cryptography protocols theoretically establishes the strength of a protocol and the constraints under which it can perform, it does not take into account
the overall design of the protocol. In the past model checking has been successfully applied to classical cryptography protocols to weed out design flaws which would have otherwise gone unnoticed. Quantum key distribution protocols differ from their classical counterparts, in their ability to detect the presence of an eavesdropper while exchanging the key. Although unconditional security has been proven for both BB84\cite{proof84} and B92\cite{proof92} key distribution protocols, in this paper we show that identifying an eavesdropper's presence is constrained on the number of qubits exchanged. We first model the protocols in Communicating Quantum Processes (CQP)\cite{gay}\cite{davidson} and then explain the mechanism by which we have translated this into a PRISM model and how we analysed the protocols' capabilities.
We mainly focus on the protocols' ability to detect an active eavesdropper and the extent to which an eavesdropper can retrieve the shared key without being detected by either party. We then conclude by comparing the performance of the protocols.

\end{abstract}


%
\IEEEpeerreviewmaketitle

\section{Introduction}
Quantum cryptographic protocols have garnered much acclaim in the last two decades for their ability to provide unconditional security, which is not practically assured by 
their classical counterparts. Commercial availability of quantum infrastructure in the last decade has placed even more emphasis on developing methodologies to ascertain 
the reliability of protocols in practice. Even though, protocols are theoretically secure, our experience with classical protocols has
shown that security can be compromised during implementation. Since modelling, analysing and verifying classical protocols have worked so well, developing techniques along 
these lines seems prudent for quantum cryptographic protocols as well.

The cornerstone of quantum cryptographic protocols is the inherent probabilistic nature. Unlike classical protocols which accommodates a passive eavesdropper, wherein the 
eavesdropper can copy the bits and analyse them later, quantum protocols mandate an active eavesdropper. This constraint is promulgated by the no-cloning\cite{nocloning} theorem which 
handicaps the eavesdropper from copying qubits. To extract information from the qubits an eavesdropper will inevitably resort to measuring them in a basis which might be 
different from the encoding basis and thereby alters the state of the qubit. This action is probabilistic in nature. Moreover, quantum protocols also involve both classical
and quantum channels. Therefore we need a language that is capable of modelling probabilistic phenomenon and also takes into account both classical and quantum 
communications.

Communicating Quantum Processes (CQP)\cite{gay} is a language developed with the expert purpose of modelling quantum protocols. CQP uses the communication 
primitives of pi-calculus\cite{pi} and has capabilities for applying unitary operators, performing measurements, and a static type system that differentiates between classical and 
quantum communications. Hence CQP seems an obvious choice for modelling quantum protocols. Reasoning along the same lines, PRISM allows us to model probabilistic 
transitions, as we show later, this allows to seamlessly translate a CQP model into a PRISM model.

Previous work on analysis of BB84 by Papanikolaou \cite{papa} has reasoned about the probability of detecting an eavesdropper and corroborates the claim made by Mayers in his proof of 
unconditional security of BB84. However, this work does not model BB84 in CQP. We first model BB84 in CQP, conver the CQP model into PRSIM and check the validity of the observations made by Papanikolaou\cite{papa}.
We then proceed to show that B92's eavesdropping detection capabilities can be reasoned along the same lines.

To ensure brevity we have refrained from explaining Quantum Mechanical primitives like unitary operators, measurements and no-cloning theorem. One good resource is Nielsen and Chuang's work\cite{ic}. Also, we have only provided an elementary introduction to CQP, only to the extent to which we use it in this paper. A better and complete resource would 
be Thimothy Davidson's\cite{davidson} doctoral thesis.

\section{Preliminaries}
We are going to briefly explain quantum measurement, and working of BB84 and B92 protocols.
\subsection{Quantum Measurement}

It is inherent with any quantum mechanical system that any measurement 
done on the system will induce some irreversible disturbances. We are going to rely on
this property of qubits heavily in any quantum cryptographic protocols.\\

Any quantum system can be represented as a vector in an $n$ dimensional complex
Hilbert space. Measuring this quantum system can only give a set of priviliged results
namely those associated with the basis vectors of the state space.\\
For example, consider a \textit{2-dimensional} complex Hilbert spcae with $\Ket{0}$ and $\Ket{1}$
as basis vectors. Lets say the vector $\Ket{\psi} = \alpha.\Ket{0} + \beta.\Ket{1}$
describes the system. If we try to measure the system in the basis $\{0,1\}$, then the system
changes to a new state, either $\Ket{\psi'} = \Ket{0}$ or $\Ket{\psi'} = \Ket{1}$
permanently. It has a probability ${|\alpha|}^2$ of changing into 
$\Ket{\psi'} = \Ket{0}$ and a probability $|\beta|^2$ of 
changing into $\Ket{\psi'} = \Ket{1}$. 
Also, ${|\alpha|}^2 + {|\beta|}^2 = 1$. We can also measure the system
in whichever basis that we choose. Lets measure the system in another
basis $\{+,-\}$, where \\ $\Ket{+} = \frac{1}{\sqrt{2}}(\Ket{0}+\Ket{1}$
and $\Ket{-} = \frac{1}{\sqrt{2}}(\Ket{0}-\Ket{1}$, then the quantum
state can be represented as 
$\Ket{\psi} = \frac{(\alpha+\beta)}{\sqrt{2}}(\Ket{+})+
\frac{(\alpha-\beta)}{\sqrt{2}}(\Ket{-})$.\\
Measuring this system in the basis $\{+,-\}$ will yield $\Ket{+}$ and $\Ket{-}$ with
probability $\frac{(\alpha+\beta)^2}{2}$ and $\frac{(\alpha-\beta)^2}{2}$ 
respectively. 

\subsection{BB84 QKD protocol}
A and B want to establish a secret for secure communication. A sends the encoding
of some bits in the \rec,\dia basis to B on the quantum channel. 
B then chooses a random sequences of bases and measures the qubit sent
by A in that basis. If the basis of Alice and Bob are equal then the
B obtains the classical bit chosen by Alice other wise she randomly
gets $\{0,1\}$. A and B then use the classical channel to exchange the basis and the
corresponding measurements of qubits to decide upon a shared key or to detect the presence
of an eavesdropper.

\subsection{Understanding B92}
Unlike BB84 where each classical bit has two different encoding depending
on the basis used, B92 has only one. In other words there is a one to one correspondence
between the classical bits and qubits exchanged. If Alice wants to send a classical bit
$0$ to Bob she sends \recnot and if she wants to send $1$ she sends \diaone.
The rest of the steps involved are the same as in BB84.

\subsection{Eavesdropping Attacker}
As mentioned earlier, whenever Eve measures the qubits that
are in transit to Bob from Alice, she makes a permanent change to the state
of qubits if she doesn't use the same basis as that of Alice. In BB84 
protocol if on some qubits both Alice and Bob use the same basis to 
encode and measure but Bob decodes a classical bit different 
from what Alice encoded, suggests the presence of Eve.
In B92 as well, Alice and Bob should obtain the opposite results when the encoding basis is
the same, then an attacker is present. We are
assuming the qubit channel shared by all the participants noiseless.

\section{Formalising in CQP}
A brief overview of CQP calculus is provided and then we proceed to formalise both the protocols in CQP. An example of BB84-Bit Commitment Protocol in CQP\cite{gay} was
give by Simon and Gay and our formalisation uses the same techniques.

A protocol at any given point of time has multiple participants, like \textit{Alice} and \textit{Bob} which are legitimate entities involved and also
adversaries like \text{Eve}. These entities are collectively known as \textit{agents}. Agents communicate with each other via communication channels to exchange
information. The working of the agents is encapsulated by \textit{processes}. Every agent has more than one process, and at any given time its possible that
more than one process is in action. These processes can be reasonably thought of as \textit{states} in finite state automatons and every process transitions to another
or terminates. CQP allows us to impose a probabilistic distribution across these transitions. Also processes in CQP can be parametrised.
\\
\begin{enumerate}
    \item channels are declared by the  \latinword{new} keyword. \\ For example to declare a new qubit channel, we write (\latinword{new} qubitChannel:\textasciicircum[\latinword{Qbit}]),
          where \latinword{Qbit} is the data type qubitChannel is constrained to and "\textasciicircum" identifies it as a channel.
    \item variables can be declared within a process like so, (\latinword{qbit} q).
    \item \textit{Process Output:} ${c![x].P_{i+1}}$ to send the data stored by variable \textit{x} along channel \textit{c} and then proceed with process ${P_{i+1}}$. 
    \item \textit{Process Input:} ${c?[x].P_{i+1}}$ to receive along channel \textit{c} and then proceed with process ${P_{i+1}}$.
    \item \textit{Process action:} ${e}.P_{i+1}$ evaluates expression \textit{e} and then proceeds with process ${P_{i+1}}$
    \item \textit{Process decision:} ${\latinword{if} e \latinword{then} P_{i+1} \latinword{else} P_{i+2}}$ if the expression $e$ evaluates to \latinword{true} then proceed with process $P_{i+1}$ else $P_{i+2}$
    \item \textit{Terminate:} ${P_{i}.\mathbf{0}}$ the process terminates after $P_i$. 
\end{enumerate}

\subsection{Formalising BB84}
We identify that \textit{Alice}, \textit{Bob} are the primary agents of the protocol and to analyse the effects of an eavesdropper \textit{Eve} becomes an
agent of the system as well. As described above channels can only transport messages of a particular type. We have \latinword{qubitChannel} to transport qubits,
\latinword{intChannel} for integers and \latinword{decisionChannel}, \latinword{decisionFlagChannel}, \latinword{randomBitChannel} for bits. Technically one bit channel would suffice.

However having two different channels that are used at two different stages in the protocol helps us to convert the CQP-model into PRISM as will be elaborated
in the next section. We have also made use of \latinword{List} type, with its associated functions of \latinword{hd}, \latinword{tl}, \latinword{[]} and \latinword{@} for 
reading the first element, dropping the first element, an empty list and placing data at the tail of the list respectively. The use of these functions is demonstrated by Gay et al.\cite{gay}.

\begin{itemize}
    \item \textit{System} is parameterized by a \latinword{bitList}, which constitutes the classical (see Figure 1).
          bits that need to be exchanged between \textit{Alice} and \textit{Bob}
    \item \textit{Random} agent creates a random bit and sends it via the                           \latinword{radomBitChannel}
    \item \textit{Alice} first sends the length of the number of bits to be exchanged
          with \textit{Bob}, i.e the length of \latinword{bitList}.
    \item Upon sending the length of the bit list, \textit{Alice} continues with the 
          process \textit{AliceSend}. This is a recursive process which terminates after
          sending all the bits in \latinword{bitList}. \textit{AliceSend} first receives a
          random bit from \latinword{randomBitChannel}, if the value received is equal to zero then the \latinword{qubit q} is encoded in the rectilinear basis else it is encoded
          in the diagonal basis. \latinword{(qubit q)} creates a new qubit \latinword{q} 
          initialised to $\Ket{0}$. Hence an operation of $X$ on $q$ to create $\Ket{1}$ and
          $X$ or  $X,H$ to convert it into $\Ket{+}$ and $\Ket{-}$ respectively. 
          \textit{AliceSend} then sends the qubit $q$ via \latinword{qubitChannel} to be 
          received by \textit{Bob}. The random bits are stores in \latinword{encodeBitList} to be used later when both the entities decide upon the key.
    \item \textit{Bob} receives the length of the \latinword{bitList} and then continues with
          \textit{BobReceive} process. Like \textit{AliceSend}, this is a recursive process
          which terminates after receiving all the bits. \textit{BobReceive} then uses a 
          random bit from \latinword{randomBitChannel}, if this bit is zero then \textit{Bob} 
          measures the received qubit in the rectilinear basis else in the diagonal basis.
          We used a list that stores a couplet, where we store the random bit and the 
          corresponding measurement.
    \item After exchanging the qubits, \textit{Alice} and \textit{Bob} continue with 
          \latinword{AliceReveal} and \latinword{BobFinal} respectively. 
          \latinword{AliceReveal} sends the basis that she used for encoding
          via the \latinword{decisionBitChannel}. \latinword{BobFinal} upon receiving
          this basis elements checks whether the basis he measured in the same as of
          that of \textit{Alice} in which case, he sends an acknowledgement via
          \latinword{decisionFlagChannel} to \textit{Alice} and the corresponding bit
          he measured. \textit{Alice} checks if the measurement that \textit{Bob} made is
          the same as that of the intended bit. Since we are dealing with channels without
          any noise, if the measurement \textit{Bob} made does not match, \textit{Alice} 
          straight away confirms the presence of an attacker and sends an \latinword{eveDetect} flag to \textit{Bob}. 
\end{itemize}

\subsection{Formalising B92}
Since $B92$ and $BB84$ only differ in how they encode the qubits, we can modify the CQP
formalisation of $BB84$ for $B92$ (see Figure 2). \textit{AliceSend} does not encode
the qubit in a random basis. If the \latinword{bitList} element is equal to zero then she
sends $\Ket{0}$ else if the element in equal to one then $\Ket{+}$ is exchanged. With few
modifications to \textit{AliceSend} in BB84, we can adopt it model B92. These modifications
are presented in \textit{Figure 2}.

\section{Modelling and Analysis in PRISM}
Conversion from CQP to PRISM is a step by step process. This conversion for a subset of commands has been done by
Ware in his Master's thesis\cite{ware}. We are going to use the same procedure (See Appendix for the PRISM models). 
In the previous section we have mentioned that we have used \latinword{List}
type. Unfortunately a parallel for this type does not exist for \latinword{PRISM}. To overcome this handicap we will
have to modify the model, in both the protocols the public discussion starts after both the parties have exchanged
all the qubits. Instead in the PRISM model after every qubit exchange, both the parties proceed to exchange the
encoding basis and measured bit to establish the validity of the qubit. This way we can ensure that the original
characteristics of the protocol remain intact.

\begin{itemize}
    \item all the channels in the CQP model are defined as global variables in the \latinword{PRISM} model.
    \item the \latinword{PRISM} model constitutes of three modules representing the different agents in the CQP modelling
    \item on the \latinword{qbitChannel} the messages to be exchanged are limited to $[0..3]$ with $0$ representing
          $\Ket{0}$, $1$ for $\Ket{1}$, $2$ for $\Ket{+}$ and $3$ for $\Ket{-}$.
    \item when Eve is detected, both \textit{Alice} and \textit{Bob} cease to exchange any more qubits and reach their
           end state.
    \item like in the CQP model we do not create a module for Random, rather all the parties create their own random bits
          either zero or one with equal probability.
    \item after choosing a random basis to measure in there is a one-fourth probability of any of the four outcomes.
    \item the number of bits to be exchanged is set by $N$ the global variable. We check the properties of the model
          by varying the value of $N$. \textit{Alice} and \textit{Bob} iterate constrained by the value $N$ and are
          synchronised by the label \latinword{loop}.
    \item \textit{Alice} and \textit{Bob} modules terminate either after exchanging $N$ qubits or after detecting
          \textit{Eve} and are synchronised by \latinword{stop}.
\end{itemize}

\onecolumn
\includegraphics[width=\textwidth]{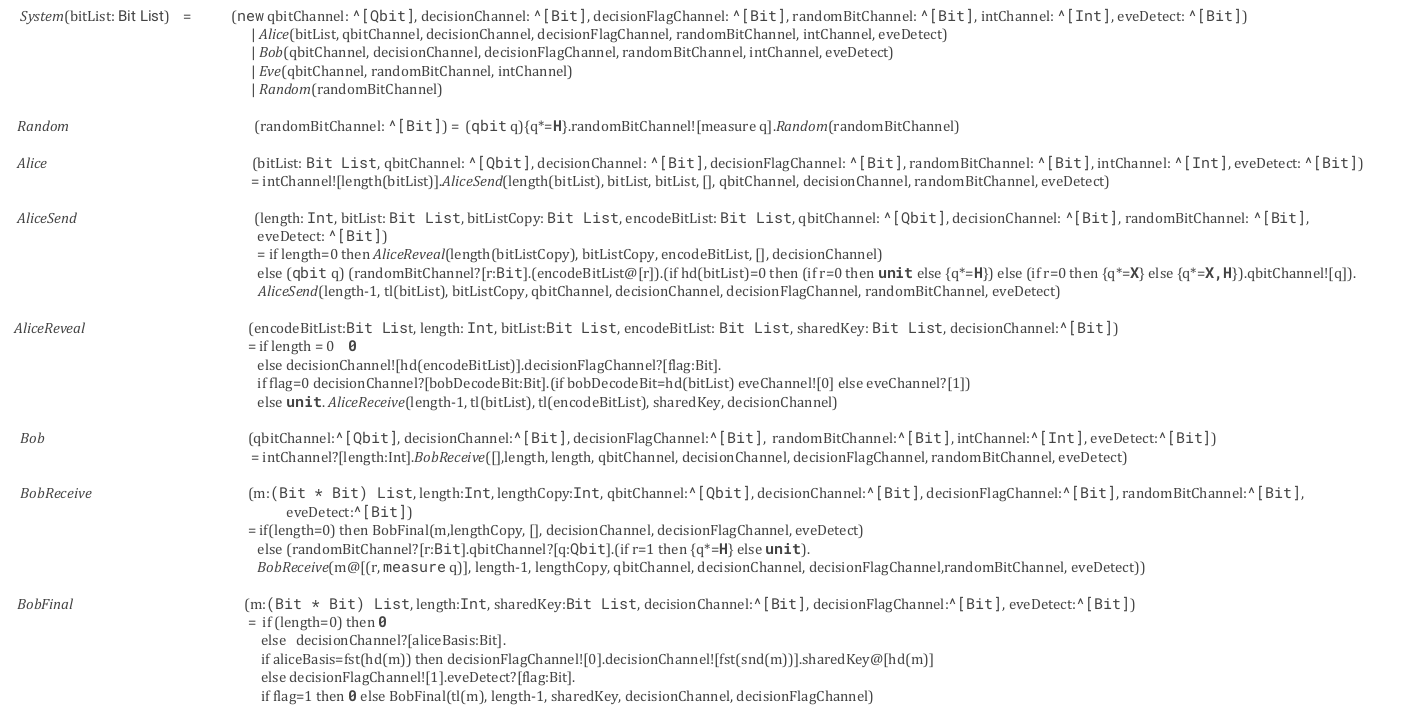}
\includegraphics[width=\textwidth]{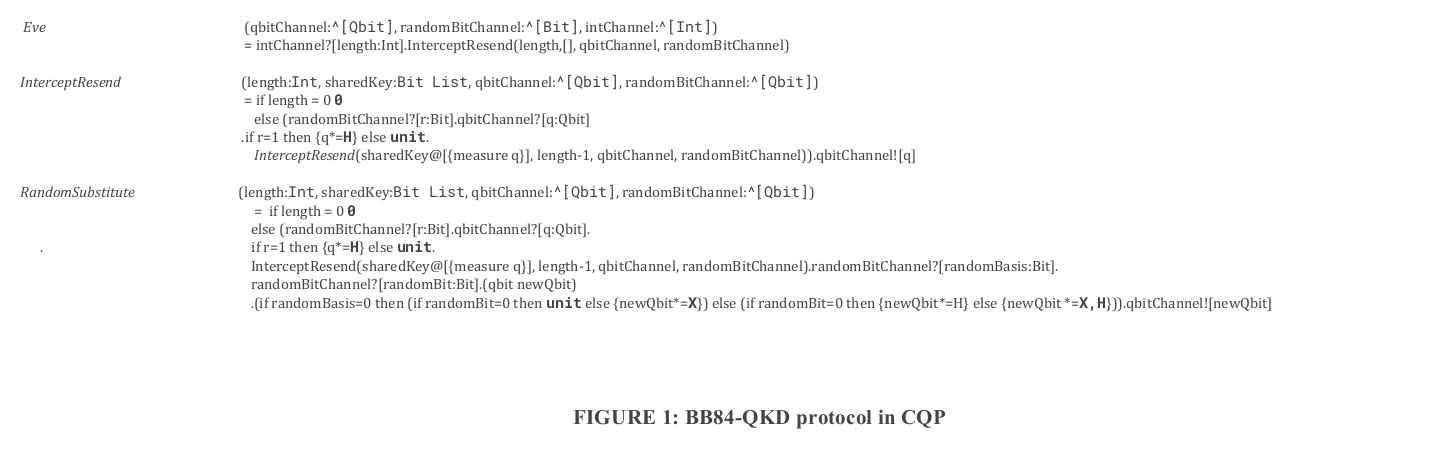}
\includegraphics[width=\textwidth]{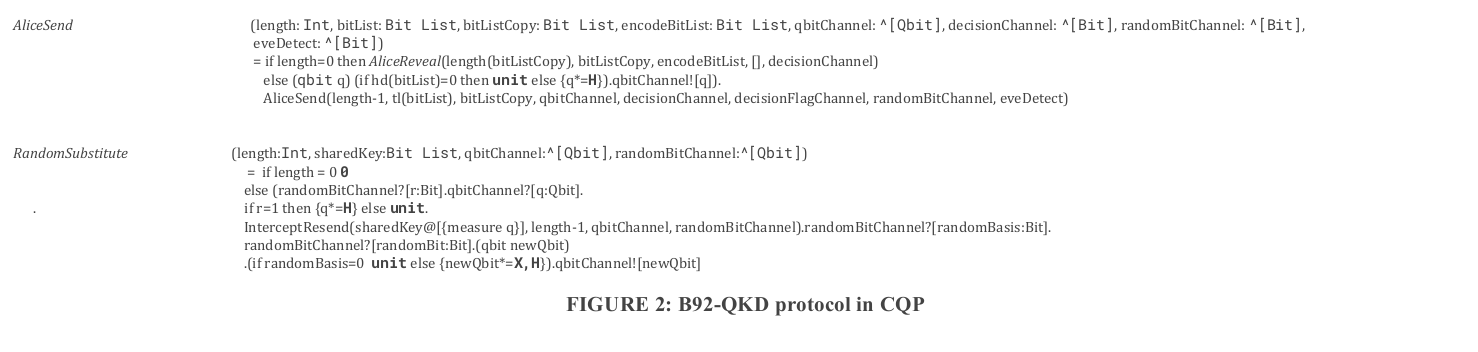}
\twocolumn
\subsection{Analysis of BB84}

With the models we have made in \latinword{PRISM} we are going to show there is a non zero probability
with which the eavesdropper can be detected and how this probability varies with the number of photons exchanged.
\newline
\latinword{PRISM} is capable of calculating probabilities of the form $P_{\sigma,\Phi} = Pr\{\sigma \models \mathbf{\Phi}\}$, i.e, given a \latinword{PRISM}
model $\sigma$, we can calculate the probability with which the property $\mathbf{\Phi}$ holds. $\mathbf{\Phi}$ is expressed
in \textbf{PCTL}. We have two models $\sigma_1$ and $\sigma_2$ for \textit{random-substitution} and \textit{intercept-resend}, 
respectively. Both these models are parametrised by $N$ the number of qubits that both the parties exchange. 
\newline
\newline
Let $P_{ED}^n = Pr\{\sigma_n(N) \models \mathbf{\Phi_1}\}$ for $n\!\in\{1,2\}$, for the probability of eavesdropper detection and
$P_{CM}^n = Pr\{\sigma_n(N) \models \mathbf{\Phi_2}\}$ for $n\!\in\{1,2\}$ for the probability of the eavesdropper making correct measurements
for more than half of the qubits. $n=1$ for \textit{random-substitution} and $n=2$ for intercept resend. We also have $N\in[1,20]$, i.e, we start to find these probabilities starting from one qubit being exchanged
to twenty.
\newline

$\mathbf{\Phi_1}$ and $\mathbf{\Phi_2}$ are to be expressed in PCTL. $\mathbf{\Phi_1}$ is the PCTL formula corresponding to when the eavesdropper is 
detected. From the PRISM model for BB84 (in \textit{Appendix A}), whenever an eavesdropper is detected \textit{Alice} is in \textit{aliceState=15}
and \textit{Bob} is in state \textit{bobState=10}. The corresponding expression for $\mathbf{\Phi_1}$ and their property expression in PRISM:

\begin{center}
$\mathbf{\Phi_1} = \{(aliceState=15) \wedge (bobState=10)\}$
\latinword{P=?[F(aliceState=15)\&(bobState=10)]}
\end{center}

Similarly for $\mathbf{\Phi_2}$ which gives the probability of eavesdropper measuring more than half of the exchanged qubits correctly is
\begin{center}
$\mathbf{\Phi_2} = \textbf{true}$  $\mathcal{U}$  $(correctMeasurement>N/2)$ 
\latinword{P=?[F(correctMeasurement$>\frac{N}{2}$)]}
\end{center}

\setcounter{figure}{3}
\begin{figure}[h!]
\centering
\includegraphics[width =80mm, scale=1.0]{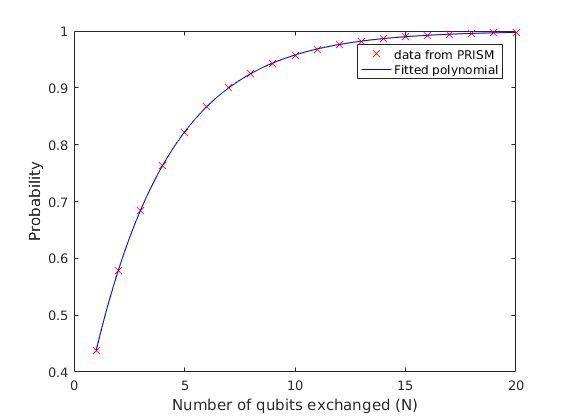}
\caption{Probability of detecting eavesdropper for BB84 Random Substitution}
\end{figure}

\begin{figure}[h!]
\centering
\includegraphics[width =80mm, scale=1.0]{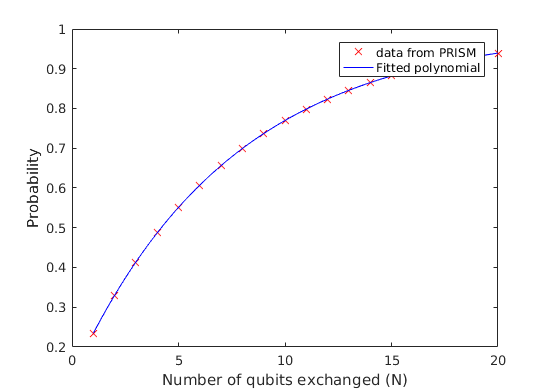}
\caption{Probability of detecting eavesdropper for BB84 Intercept Resend}
\end{figure}

\begin{table}[h!]
\centering
\begin{tabular}{|c|c|c|}
\hline
N & $P_{ED}^1$ & $P_{ED}^2$ \\
\hline 
5  &  0.822     & 0.5512    \\
10 &  0.9577    & 0.7698   \\
15 &  0.9899    & 0.8819   \\
20 &  0.9976    & 0.9394   \\
\hline
\end{tabular}
\caption{Probability of detecting eavesdropper for BB84-QKD}
\end{table}

\begin{figure}[h!]
\centering
\includegraphics[width =80mm, scale=1.0]{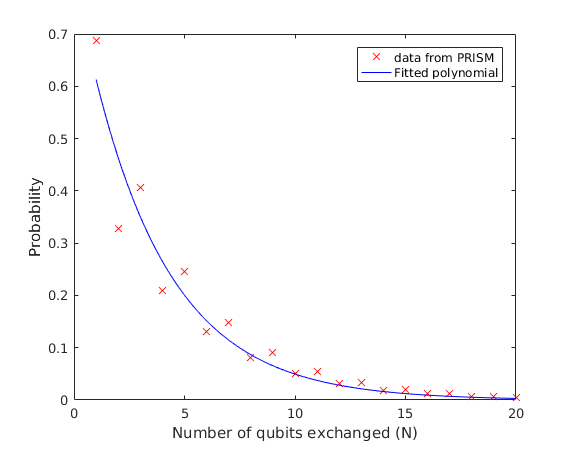}
\caption{Probability of measuring more than $\frac{N}{2}$ qubits correctly for BB84 Random Substitution}
\end{figure}

\begin{figure}[h!]
\centering
\includegraphics[width =80mm, scale=1.0]{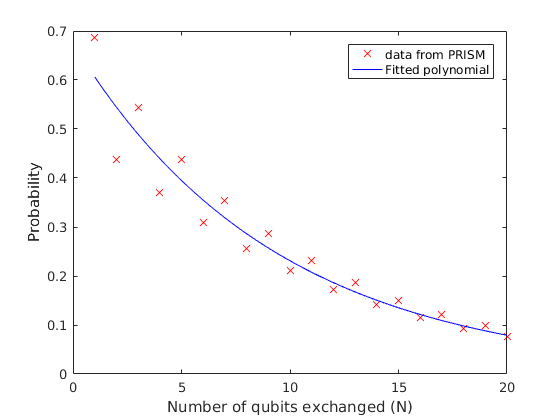}
\caption{Probability of measuring more than $\frac{N}{2}$ qubits correctly for BB84 Intercept Resend}
\end{figure}

\begin{table}[h!]
\centering
\begin{tabular}{|c|c|c|}
\hline
N & $P_{CM}^1$ & $P_{CM}^2$ \\
\hline 
5 & 0.2458    & 0.4370       \\
10 & 0.0501   & 0.2111   \\
15 & 0.0188   & 0.1510   \\
20 & 0.00425  & 0.0756   \\
\hline
\end{tabular}
\caption{Probability of eavesdropper measuring more than half of the qubits correct for BB84-QKD}
\end{table}

TABLE I and TABLE II have probabilities that are observed from \latinword{PRISM}. Using the Curve Fitting tool of MATLAB, and using the 
Marquardt-Levenberg nonlinear least squares algorithm for curve-fitting we have come up with the equation that best fits these probabilities.

We observed that \newline
\begin{center}
$P_{ED}^1 \approx Pr\{\sigma_1(N) \models \mathbf{\Phi_1}\} = (1-(0.75)e^{-0.2877N})$
$P_{ED}^2 \approx Pr\{\sigma_2(N) \models \mathbf{\Phi_1}\} = (1-(0.8750)e^{-0.1335N})$
\end{center}
and
\begin{center}
$P_{CM}^1 \approx Pr\{\sigma_1(N) \models \mathbf{\Phi_2}\} = (0.8108)e^{-0.2795N}$
$P_{CM}^2 \approx Pr\{\sigma_2(N) \models \mathbf{\Phi_2}\} = (0.6750)e^{-0.1072N}$
\end{center}

Since $P_{ED}^1 > P_{ED}^2$, the probability of eavesdropper getting detected is higher when the eavesdropper resorts to random-substitution.

Also it has to be noted that:
$$\lim_{N\to\infty} P_{ED}^1 = \lim_{N\to\infty} P_{ED}^2 = 1$$ which suggests as the number of qubits exchanged increases so does the chances of detecting an eavesdropper.
$$\lim_{N\to\infty} P_{CM}^1 = \lim_{N\to\infty} P_{CM}^2 = 0$$ reaffirms the theoretical results obtained by Mayers\cite{proof84}, wherein he states \newline \textit{"amount of Shannon's information available to Eve must decrease exponentially
fast as N increases."}.
Both these observations reaffirm the results obtained by Papanikolaou\cite{papa}.

\subsection{Analysis of B92}
We use the same notations as in the previous subsection. The only change being the PCTL expressions.
Referring to PRISM model for B92(\textit{Appendix B}), eavesdropper is detected when \textit{aliceState=11} and \textit{bobState=10}.

\begin{center}
$\Phi_1 = \{(aliceState=11) \wedge (bobState=10)\}$
\latinword{P=?[F(aliceState=15)\&(bobState=10)]}
\\
$\mathbf{\phi_2} = \textbf{true}$  $\mathcal{U}$  $(correctMeasurement>N/2)$ 
\latinword{P=?[F(correctMeasurement$>\frac{N}{2}$)]}
\end{center}

\begin{table}[h!]
\centering
\begin{tabular}{|c|c|c|}
\hline
N & $P_{ED}^1$ & $P_{ED}^2$ \\
\hline 
5 &  0.8665   & 0.7123     \\
10 & 0.9683    & 0.89812    \\
15 & 0.9924   & 0.96392   \\
20 & 0.9976   & 0.98722   \\
\hline
\end{tabular}
\caption{Probability of detecting eavesdropper for B92-QKD}
\end{table}

\begin{figure}[h!]
\centering
\includegraphics[width =80mm, scale=1.0]{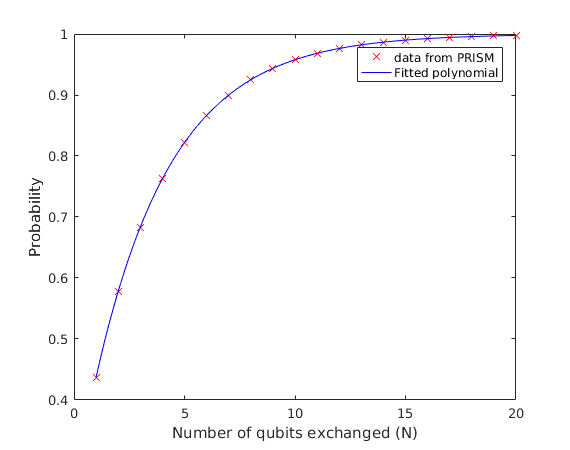}
\caption{Probability of detecting eavesdropper for B92 Random Substitution}
\end{figure}

\begin{figure}[h!]
\centering
\includegraphics[width =80mm, scale=1.0]{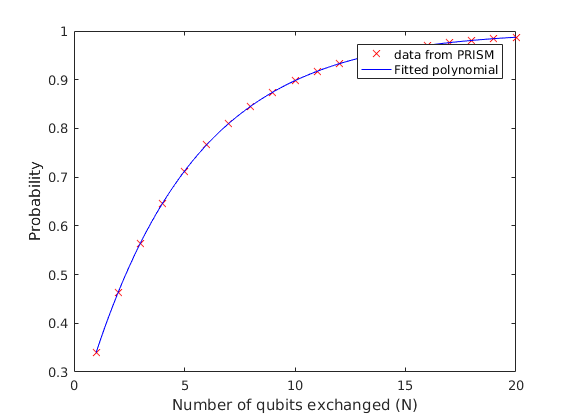}
\caption{Probability of detecting eavesdropper for B92 Intercept Resend}
\end{figure}

\begin{table}[h!]
\centering
\begin{tabular}{|c|c|c|}
\hline
N & $P_{CM}^1$ & $P_{CM}^2$ \\
\hline 
5 &  0.2458    & 0.3345      \\
10 & 0.0501    & 0.1083    \\
15 & 0.0201    & 0.0592   \\
20 & 0.0042    & 0.0199   \\
\hline
\end{tabular}
\caption{Probability of eavesdropper measuring more than half of the qubits correct for BB84-QKD}
\end{table}

\begin{figure}[h!]
\centering
\includegraphics[width =80mm, scale=1.0]{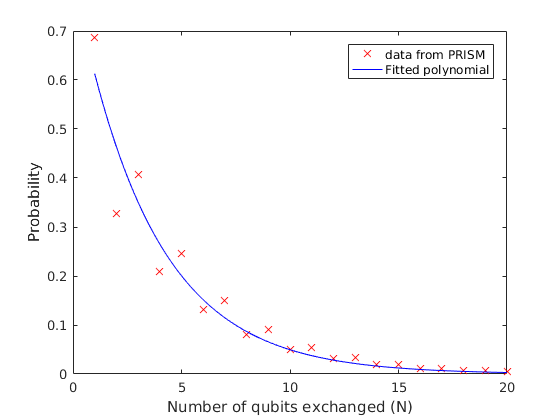}
\caption{Probability of measuring more than $\frac{N}{2}$ qubits correctly for B92 Random Substitution}
\end{figure}

\begin{figure}[h!]
\centering
\includegraphics[width =80mm, scale=1.0]{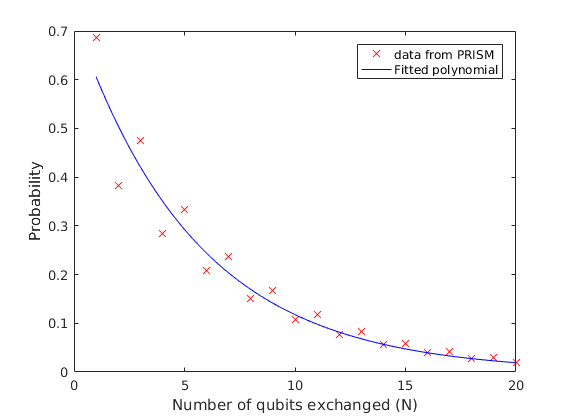}
\caption{Probability of measuring more than $\frac{N}{2}$ qubits correctly for B92 Intercept Resend}
\end{figure}

After using the curve fitting algorithm to approximate the results to an equation we have:
\begin{center}
$P_{ED}^1 \approx Pr\{\sigma_1(N) \models \mathbf{\Phi_1}\} = (1-(0.75)e^{-0.2877N})$
$P_{ED}^2 \approx Pr\{\sigma_2(N) \models \mathbf{\Phi_1}\} = (1-(0.8125)e^{-0.2795N})$
\end{center}
and
\begin{center}
$P_{CM}^1 \approx Pr\{\sigma_1(N) \models \mathbf{\Phi_2}\} = (0.8108)e^{-0.2795N}$
$P_{CM}^2 \approx Pr\{\sigma_2(N) \models \mathbf{\Phi_2}\} = (0.7272)e^{-0.1821N}$
\end{center}

We make the following obeservations:
$$\lim_{N\to\infty} P_{ED}^1 = \lim_{N\to\infty} P_{ED}^2 = 1$$.
$$\lim_{N\to\infty} P_{CM}^1 = \lim_{N\to\infty} P_{CM}^2 = 0$$.

Like the inferences made for BB84, the chances of detecting an eavesdropper increases with 
the number of qubits exchanged
and also the number of correct measurements that an eavesdropper can make decreases 
exponentially with the number of
qubits exchanged.
But unlike in BB84, for B92 we have $P_{ED}^1 < P_{ED}^2$, hence the probability of 
eavesdropper detection is higher during intercept-resend than in random substitution.

\subsection{Comparison between BB84 and B92}
Quite strangely we observe that with respect to random substitution type of attack,
both the protocols perform identically. This is substantiated by the equations
$$P_{CM}^1 \approx Pr\{\sigma_1(N) \models \mathbf{\Phi_2}\} = (0.8108)e^{-0.2795N}$$ and
$$P_{ED}^1 \approx Pr\{\sigma_1(N) \models \mathbf{\Phi_1}\} = (1-(0.75)e^{-0.2877N})$$
However with respect to intercept resend style attacks they differ markedly, as 
evidenced by Fig. 12 and Fig. 13.   

\begin{figure}[h!]
\centering
\includegraphics[width =80mm, scale=1.0]{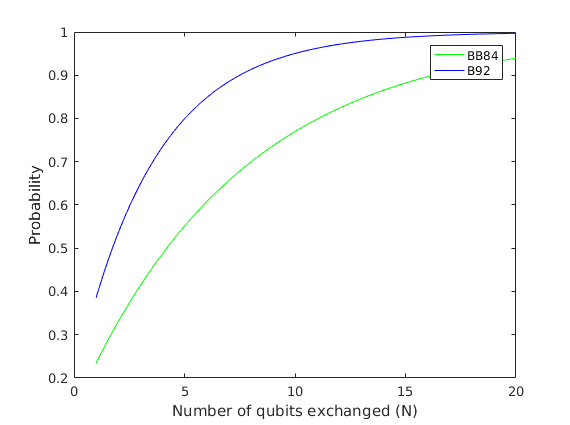}
\caption{BB84 and B92 Comparison for Intercept Resend eavesdropper detection}
\end{figure}

B92 performs better in terms of eavesdropper detection as the probability approaches unity faster than B92 and in terms of decreased number of correct measurements that can be made by
the eavesdropper.

\begin{figure}[h!]
\centering
\includegraphics[width =80mm, scale=1.0]{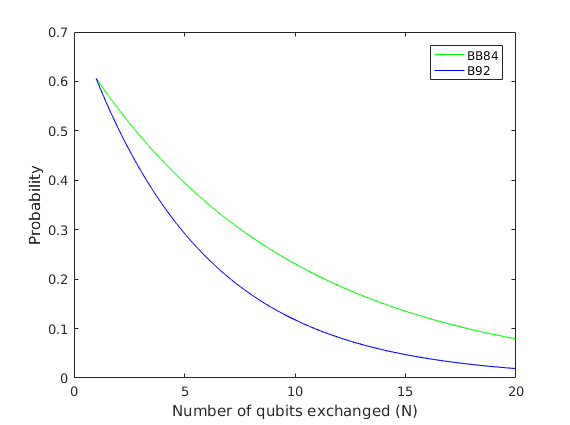}
\caption{BB84 and B92 comparison for Intercept Resend correct measurements by eavesdropper}
\end{figure}

\section{Conclusion}
We have successfully modelled BB84 protocol in CQP, showed the process  
in which we have created PRISM models from the CQP models and analysed the properties
using PCTL. We also corroborate the observations made in earlier research with our analysis. We then extended the 
technique to B92-QKD protocol and compare the performance of the two.
We infer that B92 is more resilient against an eavesdropper, with its ability to take
fewer qubits than BB84 in identifying an eavesdropper and then potentially reducing the number of correct measurements 
the eavesdropper can make.


%

\newpage

\appendices
\onecolumn
\section{Model for BB84-QKD}
\includegraphics[width=\textwidth]{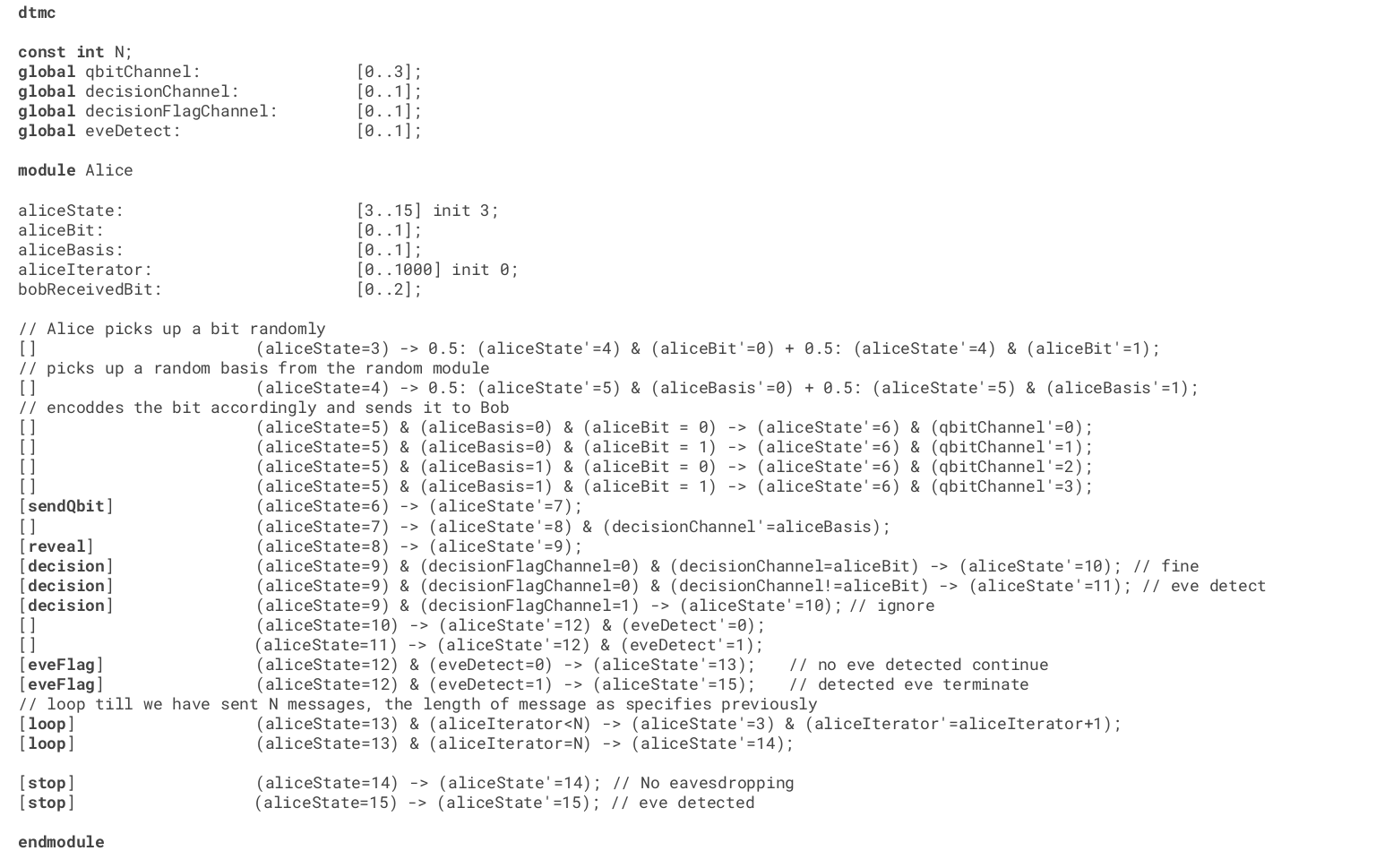}
\includegraphics[width=\textwidth]{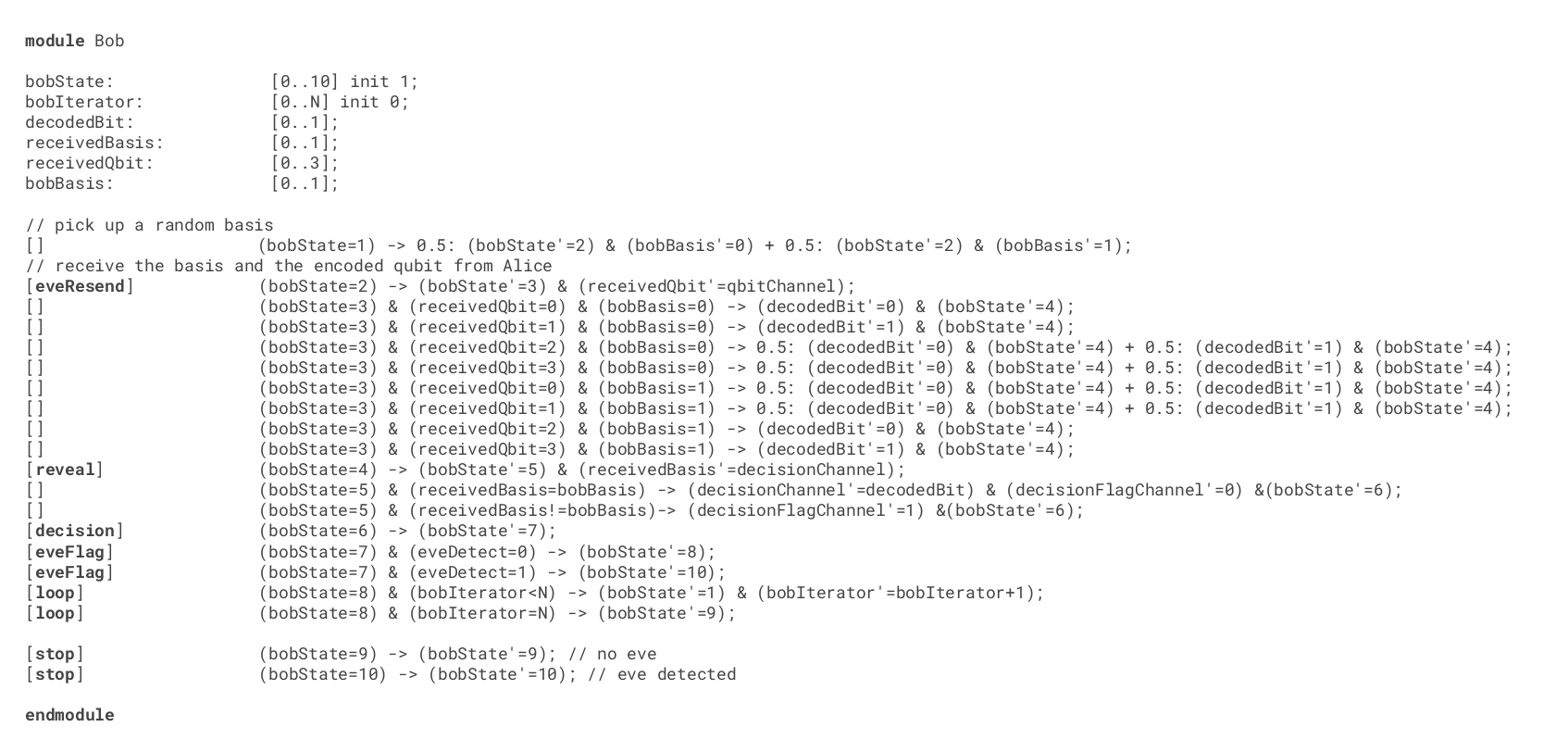}
\includegraphics[width=\textwidth]{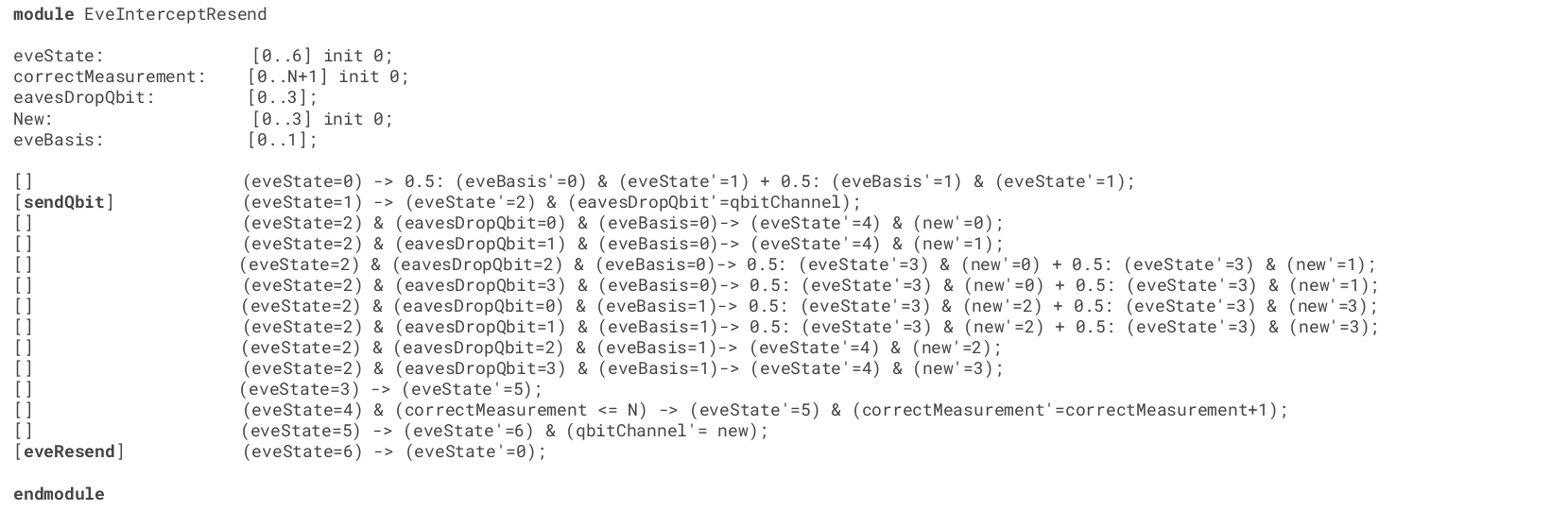}
\includegraphics[width=\textwidth]{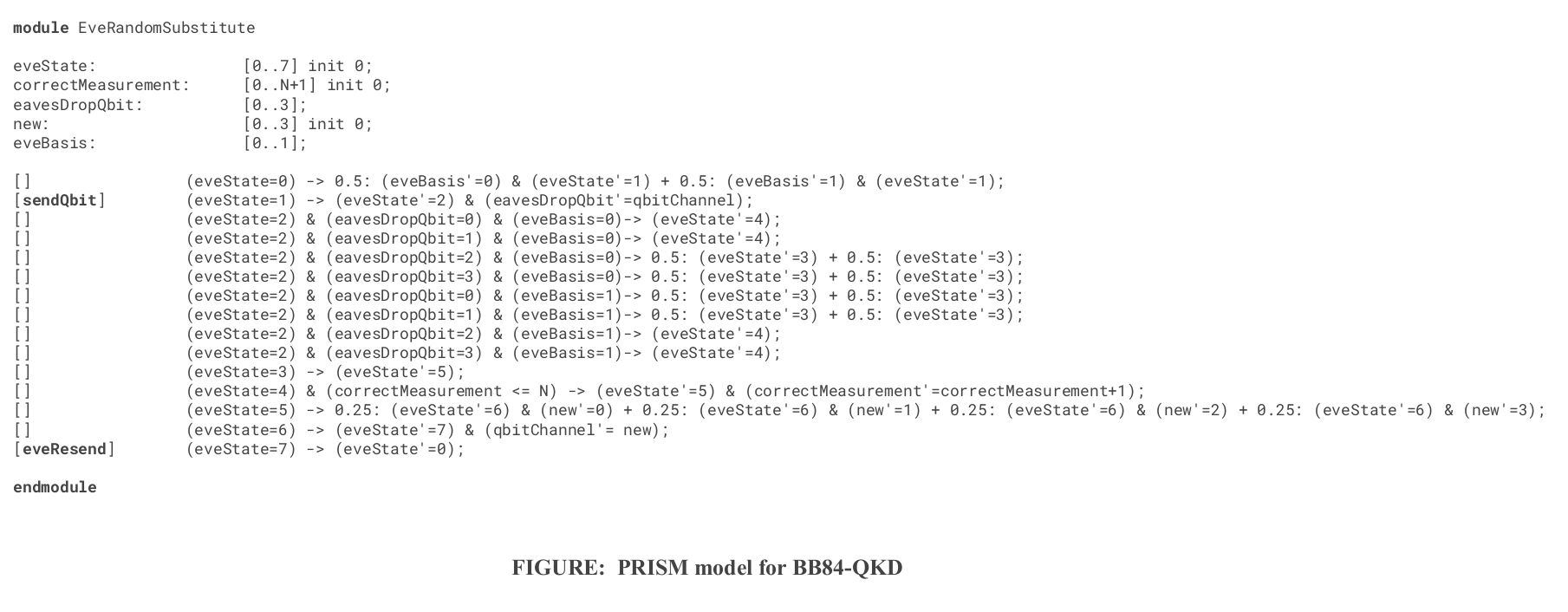}
\section{Model for B92-QKD}
\includegraphics[width=\textwidth]{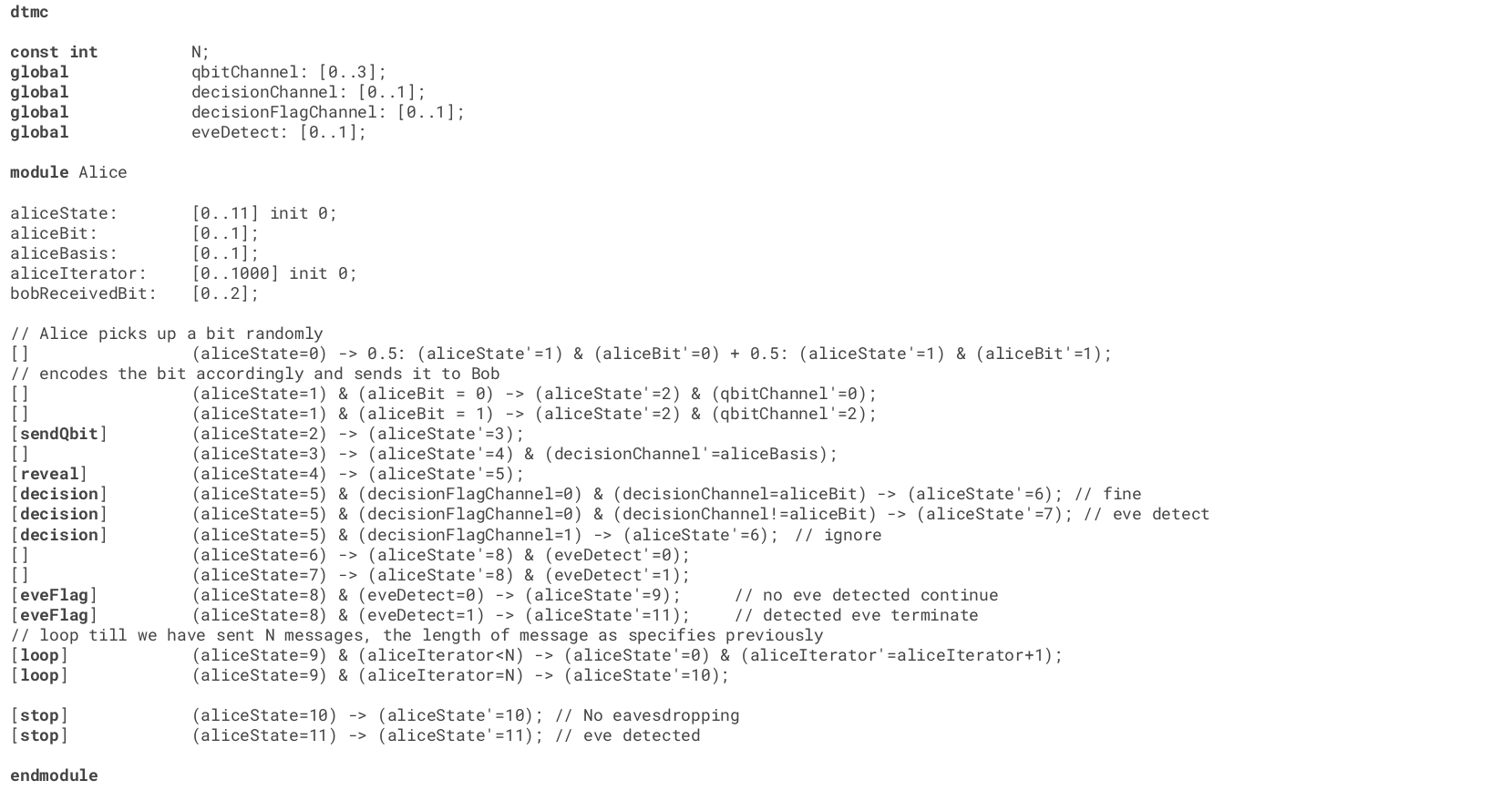}
\includegraphics[width=\textwidth]{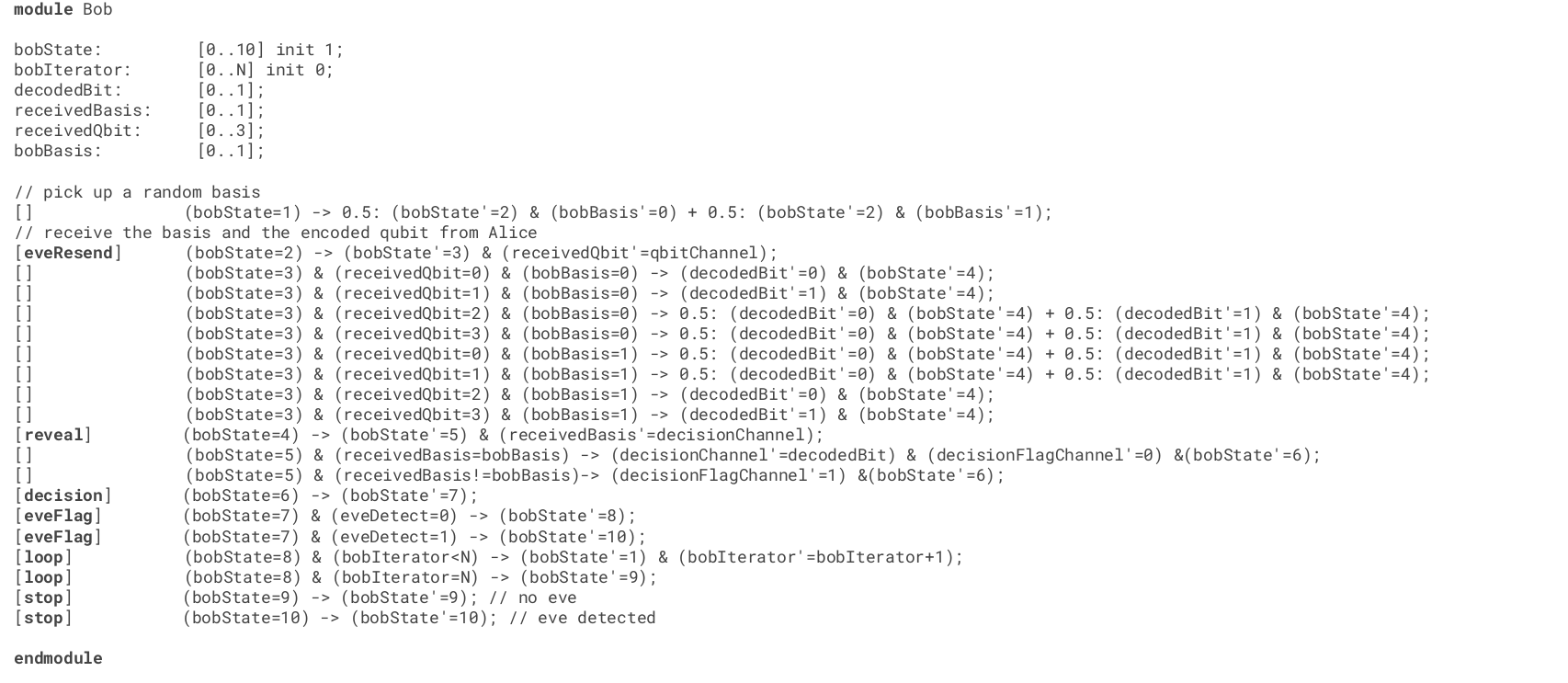}
\includegraphics[width=\textwidth]{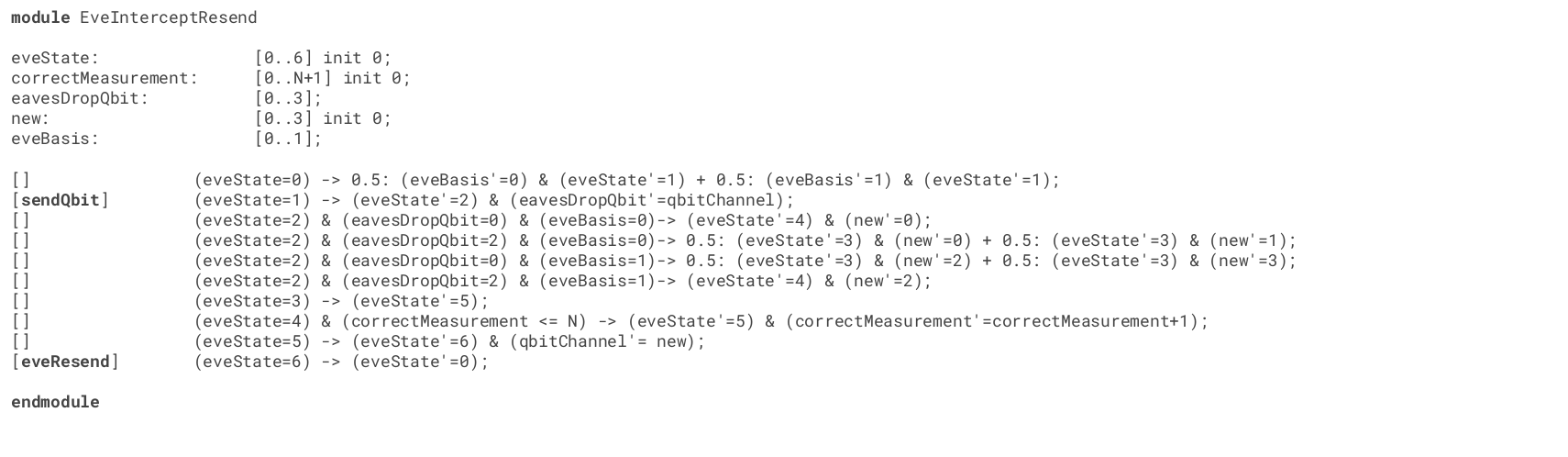}
\includegraphics[width=\textwidth]{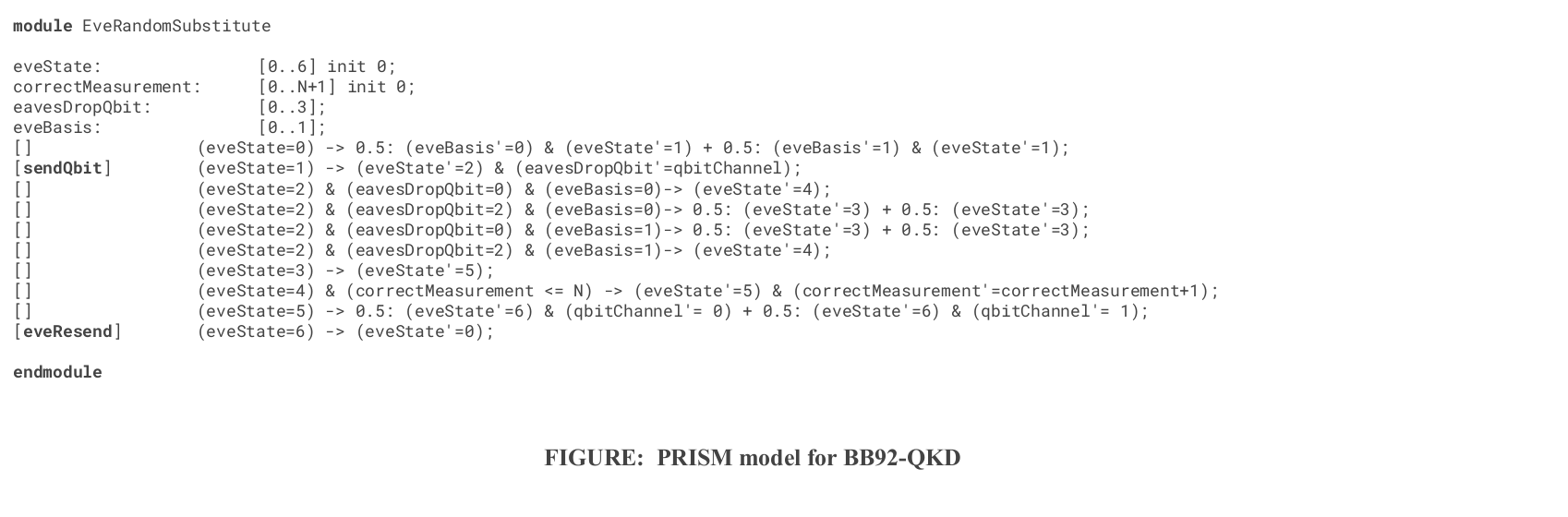}
\twocolumn

\end{document}